\documentclass{IEEEcsmag}

\usepackage[colorlinks,urlcolor=blue,linkcolor=blue,citecolor=blue]{hyperref}

\usepackage{graphicx}
\usepackage{adjustbox}
\usepackage{booktabs}
\usepackage{pifont}
\usepackage{amssymb}
\usepackage{verbatim}
\usepackage{bibunits}
\usepackage{tcolorbox}
\usepackage{hyperref}

\usepackage{caption}
\usepackage{subcaption}

\usepackage{color}
\usepackage{amssymb}
\usepackage{grffile}
\usepackage{relsize}
\usepackage{colortbl}
\usepackage{enumerate}
\usepackage{amsmath}
\usepackage{tikz}
\usepackage{pgf}
\usepackage{pbox}
\usepackage{rotating}
\usepackage{multirow}
\usepackage{balance}
\usepackage[english]{babel}
\usepackage{tablefootnote}
\usepackage{flushend}

\usepackage[nocompress]{cite}

\usepackage{soul}
\usepackage{upmath}

\jvol{XX}
\jnum{XX}
\paper{8}
\jmonth{November/December}
\jname{IEEE Software}
\pubyear{2023}

\begin{document}

\sptitle{Department: Head}
\editor{Editor: Name, xxxx@email}

\title{Students' Perspective on AI Code Completion: Benefits and Challenges}

\author{Wannita Takerngsaksiri}
\affil{Monash University, Australia.}

\author{Cleshan Warusavitarne}
\affil{Monash University, Australia.}

\author{Christian Yaacoub}
\affil{Monash University, Australia.}

\author{Matthew Hee Keng Hou}
\affil{Monash University, Australia.}

\author{Chakkrit (Kla) Tantithamthavorn}
\affil{Monash University, Australia.}

\markboth{Department Head}{Paper title}

\begin{abstract}
AI Code Completion (e.g., GitHub's Copilot, Amazon CodeWhisperer) has revolutionized the way in which computer science students interact with programming languages. 
However, these tools are not available for free public use, preventing us from conducting our research. 
In addition, AI code completion has been studied from developers’ perspective, not students' perspective who represent the future generation of our digital world.
In this article, we investigated the benefits, challenges, and expectations of AI code completion from students' perspectives and introduced \emph{AutoAurora}, an AI code completion tool integrated into the Visual Studio Code Extension as a research instrument.
Through an interview study with ten participants, we found that  AI code completion enhanced students' productivity and efficiency by providing correct syntax suggestions, offering alternative solutions, and functioning as a coding tutor. 
However, the over-reliance on AI code completion may lead to a surface-level understanding of programming concepts, diminishing problem-solving skills and restricting creativity. 
In the future, AI code completion must be explainable to facilitate the learning of coding concepts.


\end{abstract}

\maketitle

\section{Introduction}


In a world where technology is ever-changing and evolving, the impact of these technologies is becoming more prevalent in our society. 
Among these advancements, AI code completion is one of the major changes within the technology field.  
AI code completion is a tool in which an AI model would auto-complete the user's code whilst typing, allowing for reduced typos as well as providing users with helpful suggestions of practical code implementations based upon the contact of what was written. 
As such the introduction of these tools and technology has profoundly shaped the way computer science students learn and interact with programming languages. 

Computer science students require learning knowledge of a wide range of programming languages, algorithms, and coding syntax. 
Thus, the AI code completion tools can help students generate source code and provide promising assistance to aid in this learning process. 
There are many existing code completion tools available including GitHub's CoPilot and Amazon CodeWhisperer that assist programmers by generating code in real-time. 
However, AI code completion has been studied from developers' perspective~\cite{ernst2022ai,puryear2022github,kazemitabaar2023studying,bull2023generative},---\emph{not from students' perspective who represent the future generation of our digital world.
In addition, these tools are not available for free public use, hindering the wide adoption of such tools in classrooms.}


In this article, we investigated the benefits, challenges, and expectations when adopting AI code completion from students' perspectives. 
However, the lack of free public access to AI code completion tools hinders their widespread adoption among students.
Thus, we introduced \emph{AutoAurora}, an AI code completion tool integrated into the Visual Studio Code Extension as a research instrument.
Then, students were asked to use AI code completion to complete two programming tasks, followed by an interview.
Finally, we performed data analysis on the interview responses to develop a taxonomy of benefits, challenges, and expectations of AI code completion from students' perspectives.

\subsection{AutoAurora: A Free and Open AI Code Completion Tool for the Public.}

\begin{figure}[t]
    \centering
    \includegraphics[width=\columnwidth]{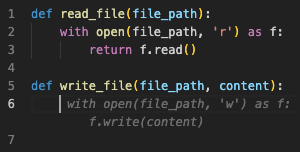}
    \caption{An example of read/write file function suggested by our AutoAurora code completion tool in the Visual Studio Code.}
    \label{fig:ex_rw}
\end{figure}

AutoAurora is developed as a Visual Studio Code extension\footnote{https://marketplace.visualstudio.com/items?itemName= PyCoder.AutoAurora}
AutoAurora is built on top of the StarCoder large language model, and is available for free public use, open-source, and easy to install.
Figure~\ref{fig:ex_rw} presents an example usage scenario of AI code completion.
The code completion model becomes active when the user prompts their code input (Lines 1-5).
Then, the code completion model will generate suggestions as the grey-colored code segments (Line 6). 
Users can then choose to accept these suggestions or continue writing code in their preferred manner.









\subsection{Key Findings}

Based on the interviews of ten participants, we draw the following key findings.

\begin{itemize}
    \item AI code completion enhanced students’ productivity and efficiency by providing correct syntax suggestions, offering alternative solutions, and functioning as a coding tutor. 
    \item However, the over-reliance on AI code completion may lead to a surface-level understanding of programming concepts, diminishing problem-solving skills and restricting creativity. 
    \item In the future, AI code completion must be explainable to facilitate the learning of coding concepts.
\end{itemize}

\hspace{4mm}

\section{AI Code Completion in a Nutshell}
\subsection{AI Code Completion in the Literature}
AI code completion is a tool designed to enhance productivity by providing real-time code suggestions and auto-completing your code as you type, all based on the current context. 
AI code completion relies on large language models (LLMs) that are trained on a large corpus of source code and natural languages (called the code completion model).

Traditionally, researchers leverage heuristic and statistical language techniques to develop a code completion model.
For example, Hou~\ea~leverages a rule-based technique to recommend code.
Robbes~\ea~\cite{robbes2008program} leverages a program history to recommend code based on historical patterns.
However, these techniques rely heavily on manually crafted rules and patterns, which are costly and time-consuming.
Hindle~\ea~\cite{hindle2016naturalness} argued that source code is natural and repetitive.
Therefore, a statistical technique like n-gram has been proposed to predict the next token of the code sequence.
Nonetheless, such statistical techniques are not suitable for code that is long and complex.
To address this limitation, deep learning techniques have been proposed for code completion.

Modern code completion approaches applied various deep learning techniques such as LSTM~\cite{li2017code}, Transformers~\cite{lu2021codexglue, takerngsaksiri2023syntax}, and Large Language Models (LLMs)~\cite{openai2023gpt4} to suggest the code based on previous context.  
In recent years, such deep learning or AI code completion has also become very popular in the software industry. 
Models trained on billions of lines of code are introduced, e.g., OpenAI GPT-3, Google Palm, Meta LLaMa, and DeepMind AlphaCode.
In 2023, OpenAI released one of the most powerful generative models, GPT-4, which is capable of generating natural language text and performing a wide range of coding tasks in real time.
These code completion models are integrated into many AI coding assistants, e.g., GitHub Copilot and Amazon CodeWhisperer.

\subsection{AI Code Completion for Education}

\emph{``The nature of learning programming will change dramatically with AI-driven development Environments (AIDEs). Whether these assistants will speed up or slow down the learning process is currently an open question."} -- stated by Ernst and Bavota~\cite{ernst2022ai}.
As AI code completion has been integrated into modern IDEs, learning programming become more convenient and challenging.
Thus, researchers have started to investigate the impact of AI code completion on education.
For example,  Puryear~\ea~\cite{puryear2022github} investigated the quality of generated code provided by GitHub Copilot in a classroom.
They found that GitHub Copilot is able to generate code with high human-graded scores ranging from 68\% to 95\% and low plagiarism scores in the introductory assignments.
Align with the previous work, Kazemitabaar~\cite{kazemitabaar2023studying} investigated the performance of novice programmers using OpenAI Codex.
They found that using Codex significantly increases the code-authoring (i.e., manually write all code) performance of completion rate by 1.15x while does not decrease the manual code-modification performance.
Bull~\ea~\cite{bull2023generative} also conducted exploratory interviews with industry professionals to understand the current practice and challenges.
They found that ``it's likely that a programmer's approach to software development will shift, moving the focus from writing typical code to working alongside generative AI assistants to design and develop code solutions''.
While much research investigated the impact of AI code completion in software engineering education, little is known about the benefits, challenges, and expectations when adopting AI code completion from students' perspectives.

\section{Research Methodology}

\subsection{Goal and Research Questions}
In this article, we aim to investigate the benefits, challenges, and expectations of AI code completion from students' perspectives. 
To achieve this goal, we formulated the following research questions:

\begin{itemize}
    \item[] \textbf{RQ1.} What are the students' benefits when adopting the AI code completion tool?
    \item[] \textbf{RQ2.} What are the students' challenges when adopting the AI code completion tool?
    \item[] \textbf{RQ3.} What are the students' expectations when adopting the AI code completion tool?
\end{itemize}

\subsection{An Interview Study}
To answer our research questions, we conducted an interview with computer science students at Monash University, Australia. 
Participants were recruited by an invitation to participate in the study.
Participants must be computer science students with at least one year of programming experience. 
To ensure that the responses we received are authentic from the actual usage, each participant will use our AutoAurora, the AI Code Completion tool, with two real-world Python programming tasks.

\begin{itemize}
    \item \textbf{Task 1} Write a program that reads a text file named ``input.txt" and perform the following tasks: Remove any leading or trailing white spaces from each line, count the total number of lines in the file, create a new file named ``output.txt" and write the modified lines to it, with each line containing its line number followed by a tab character and the modified line content. On the last line of the ``output.txt" file, write the total number of lines in the file.	
    \item \textbf{Task 2} Given an m x n matrix, return all elements of the matrix in spiral order.
\end{itemize}

The primary objective of these tasks was to familiarize participants with code completion tools and encourage them to contemplate potential enhancements that could be made to such tools.
Each participant was assigned to attempt and complete both tasks using Visual Studio Code with a time limit of 10 minutes per task.
To mitigate any potential biases, a standardized programming environment was arranged for all participants, with the AutoAurora extension enabled.
Access to online resources was not restricted throughout the duration of the tasks. 
Necessary input data and example outputs were provided, allowing participants to verify their programs.
Participants were encouraged to verbalize their thoughts as they attempt all tasks as their voice and actions will be recorded to provide insights into their thought processes and behavior while using AutoAurora.
Additionally, participants' screen usage was monitored and recorded to observe their interactions with AutoAurora during code completion.
After completing the assigned tasks, or after the allotted had expired, a post-coding survey consisting of open-ended questions was administered to gather participants' feedback on benefits, challenges, and expectations of code completion tools in the context of educational purposes. 
Our survey questions were designed as follows:

\begin{enumerate}
    \item[(Q1)] How would you rate the accuracy of the code completion tool in providing relevant suggestions while you were working on the coding questions?
    \item[(Q2)] How does code completion assist you in your programming learning journey? Can you provide specific examples of how it aids their understanding of coding concepts?
    \item[(Q3)] In what ways can code completion enhance your efficiency and productivity while writing code?

    \item[(Q4)] Are there instances where code completion might not be as beneficial for students? How can educators guide students in recognizing when to use code completion and when to rely on manual coding?

    \item[(Q5)] Looking ahead, how do you envision code completion evolving to further assist students in honing their programming skills? What improvements or additional features could be integrated into future code completion tools?
\end{enumerate}
    

\subsection{Data Analysis}
During the three weeks of the recruitment and interview campaign, ten participants were engaged in our interviews. 
The data collected undergoes reflexive thematic analysis as a qualitative method to discern patterns and identify themes.
This process encompasses open coding of participant responses, with the aim of identifying recurring themes related to the benefits and challenges associated with the use of the code completion tool, as well as potential avenues for improvement.
The data analysis process included the following steps: 

\begin{enumerate}
    \item \textbf{Open Coding}: Start with open coding, which involves breaking down the responses into smaller, meaningful units and assigning codes to these units. Then, we use inductive reasoning, letting the codes emerge naturally from the data. Then, we group related codes into categories or themes.
    \item \textbf{Constant Comparison}: We continuously compare new data to existing codes and categories as we progress through the coding process. This ensures consistency and helps us refine our coding system.
    \item \textbf{Data Saturation}: We continued coding until we reached data saturation, where we no longer found new information or themes in the data.
\end{enumerate}

Ethical considerations were followed throughout the study, ensuring participant consent, privacy, and appropriate data handling. 
Permission was obtained from the Monash University Human Research Ethics Committee (MUHREC, Project ID 38109) before conducting research interviews.
Explanatory statements and consent forms were produced to ensure that participants made informed decisions to participate and consent to data collection. 
The anonymity of participants was maintained, and the study adheres to relevant research ethics guidelines. 
The de-identified data is stored securely in a Google Drive protected by multifactor authentication.

\begin{figure*}[t]
    \centering
    \includegraphics[width=\textwidth]{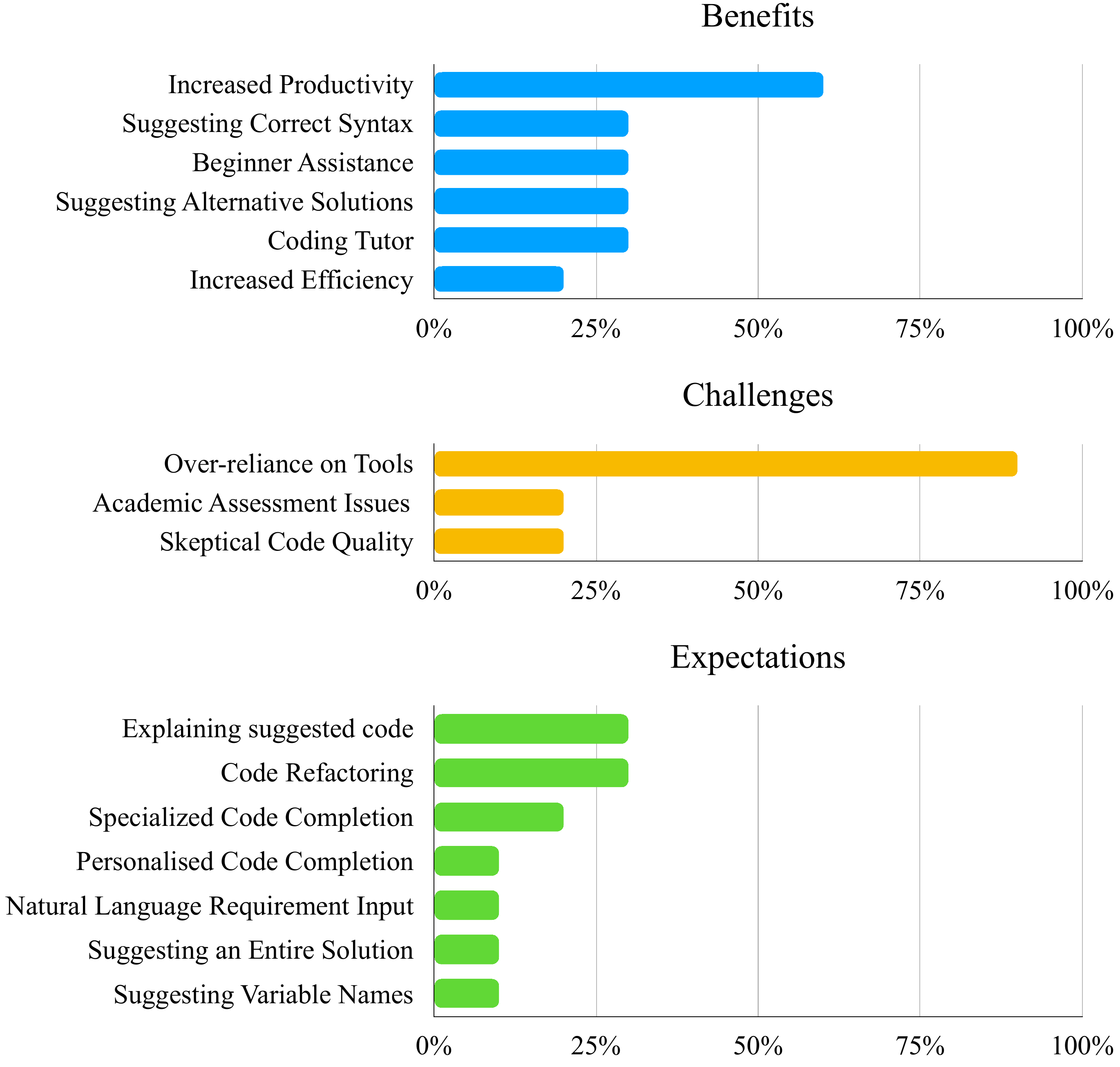}
    \caption{A summary of the benefits, challenges, and expectations of students' perspective when adopting AI code completion.}
    \label{fig:summary}
\end{figure*}

\section{Findings}

Figure~\ref{fig:summary} presents a summary of the benefits, challenges, and expectations of students' perspective when adopting AI code completion.





\section{What are the students' benefits when adopting the AI code completion tool?}


\begin{table*}[t]
\centering
\caption{The benefits of using AI code completion from students' perspectives.}
\label{table1}
\begin{adjustbox}{width=\textwidth}
\begin{tabular}{p{0.15\textwidth}|c|p{0.75\textwidth}} 
\textbf{Categories} & \textbf{\%} & \textbf{Quotes} \\
\hline
Increased Productivity & 60\% & 
\begin{tabular}[c]{@{}p{0.75\textwidth}@{}}
“It's gonna save a lot of time in the long run. In the short term. It's not gonna save that much time, but doing it every day. Definitely save a lot of time.” - P3 \\ 
“It can definitely improve efficiency and productivity. It’s not a lot for one single instance. But if you have to write a lot of files, or you have to do a lot of code work, then the time adds up”- P6\\ 
“Rather than spending a lot of time figuring out how to solve a problem it can provide a solution very quickly, rather than spending a lot of time troubleshooting.”- P8\\ 
“Most of my time is saved from typing” -P7\\ 
“It gives better access to the programming rather than self-search” -P9\\ 
“You don't have to search up every small thing you don't know about on the internet” - P10 \\
\end{tabular} \\
\hline
Suggesting \break Correct Syntax   & 30\% & 
\begin{tabular}[c]{@{}p{0.75\textwidth}@{}}
“When you're coding, you can tend to forget the syntax. You can tend to forget what frameworks you need. So things like that can take a bit of your time when you're in the flow of coding” - P1 \\ 
“I’ll say it helps learning syntax for language in terms of teaching as opposed to actually teaching programming concepts”- P3 \\ 
“Finding syntax is difficult, instead of wasting time on Google, this will be very helpful as it provides accurate formats and syntax”- P5 \\
\end{tabular} \\
\hline
Learning Assistance & 30\%  & 
\begin{tabular}[c]{@{}p{0.75\textwidth}@{}}
“It helps to understand for especially new languages, so like common things that you might miss I'd say, helps probably less so the more the deeper go into the language because doing more and more specific things, but definitely for like brevity of touching different languages, it's definitely useful”- P3\break
“It would be very beneficial in the early days of learning how to code, because initially learning to code can be very difficult as a new skill to learn because it doesn't take off from any other types of subjects or other units, it’s its own learning curve.” - P8\break
“It helps to provide a hint where I should start with and give an example of a function that responds to the problem as well so I think that’s a good start.” - P9 \\
\end{tabular} \\
\hline
Suggesting Alternative Solutions & 30\% & 
\begin{tabular}[c]{@{}p{0.75\textwidth}@{}}
“Show me a solution that i didn’t know existed”- P7\\ “it would assist in overcoming certain obstacles in assignments and other things or to give specific examples of how a problem can be solved. Also, with a current solution … it could provide an alternative way to get the same output, but in a different way.”- P8\\ “Gives insights into other methods and libraries you can use” - P10\end{tabular} \\
\hline
Coding Tutor & 30\%  & 
\begin{tabular}[c]{@{}p{0.75\textwidth}@{}}
“It can definitely serve as a guide for the students as well, like a tutor just sitting next to you and guiding you through the code” - P1\\ 
“Makes our coding faster, it reduces a lot of effort to search multiple websites for the exact code. It’s like a teacher autonomously teaching you how to code, instead of you finding many resources, and you end up spending a lot of time doing that”- P4\\ 
“it would be helpful to understand how code is structured and how a solution can be achieved when a student’s coding skills are not as advanced as they would be later in their course or years.” - P8\end{tabular} \\
\hline
Increased Efficiency & 20\% & \begin{tabular}[c]{@{}p{0.75\textwidth}@{}}
“Provides a starting point for higher level parts of what I was coding”- P6 \\ 
“Grunt work that can be handled by the tool that people otherwise don’t have to. So you can spend more time thinking about higher level solutions” - P7\\
\end{tabular} 
\end{tabular}
\end{adjustbox}
\end{table*}

\hspace{4mm}

\chapterinitial{``\emph{AI code completion enhanced students' productivity and efficiency by providing correct syntax suggestions, offering alternative solutions, and functioning as a coding tutor.}''}

Students can benefit in several ways when adopting an AI code completion tool to write code more efficiently and accurately. 
Here are some of the benefits for students:

\textbf{Increased Productivity}: AI code completion tools can help students write code faster. 
By suggesting code snippets and providing context-aware recommendations, students can reduce the time it takes to write and debug their code.
For example, a participant stated that \emph{``Most of my time is saved from typing"}.
This confirms that AI code completion will save coding time as it could reduce typing times, and not have to search for every small thing they don't know about on the Internet.

\textbf{Suggesting Correct Syntax}: Programming syntax can be challenging for students with less programming experience due to unfamiliarity (e.g., the keywords and structures used in code may be completely unfamiliar, making it difficult to understand how to write and read code).
For example, a participant stated that \emph{``It helps learning syntax for language in terms of teaching as opposed to actually teaching programming concepts''}.
This confirms that AI code completion was helpful in learning new programming languages for the first time.

\textbf{Learning Assistance}: Students found that AI code completion also provides explanations and documentation for the suggested code, helping students learn more about programming languages, syntax, and coding conventions, which is especially useful for beginners.
For example, a participant stated that \emph{``it provides a hint where I should start with and give an example of a function''}.
Therefore, such AI code completion tools help students with correct syntax, reducing the time and frustration associated with syntax errors, and allowing students to focus on learning higher-level programming concepts rather than syntax correction.

\textbf{Suggesting Alternative Solutions}: AI code completion tools can often suggest code refactoring options to improve code readability and performance. This might include suggesting more efficient algorithms or alternative code structures.
For example, a participant stated that \emph{``it shows me a solution that I didn't know existed''}.
Thus, with alternative solution suggestions, students can learn by seeing how certain tasks are implemented and gain insights into best practices.

\textbf{Coding Tutor}: Students perceive AI code completion tools as akin to having a coding tutor guide them in their coding journey.
For example, a participant stated that \emph{``like a tutor just sitting next to you and guiding you through the code''}.
This confirms that AI code completion not only assists in initial learning but also provides continuous support throughout the coding process.

\textbf{Increased Efficiency}: Apart from guiding introductory coding, AI code completion also profoundly impacts advanced coding. 
For example, a participant stated that \emph{``you can spend more time thinking about higher-level solutions''}.
By handling repetitive tasks, AI code completion enables students to focus on refining the core algorithm, elevating their concentration on more complex problem-solving aspects.

\begin{table*}[t]
\centering
\caption{The challenges of using AI code completion from students' perspectives.}
\label{table3}
\begin{adjustbox}{width=\textwidth}
\begin{tabular}{p{0.15\textwidth}|c|p{0.75\textwidth}}
\hline
\textbf{Categories}                                      & \textit{\textbf{\%}} & \textbf{Quotes} \\ \hline
Over-reliance \break on Tools & 90\% & \begin{tabular}[c]{@{}p{0.75\textwidth}@{}}“I feel like students can tend to rely on it a bit too much … I still think programmers must still be able to come up with the algorithms themselves” - P1\\ “If a student wasn’t interested in learning, he can simply complete his assignments with code completion without learning anything” - P4\\ “If students rely on this, they won’t be able to learn” - P5\\ “If you’ve got an autocomplete to help you write the algorithm, then you’re not proving that you know the algorithm, you’re just proving that you know how to use an autocomplete tool” - P6\\ “If they rely on that, they might not understand what the code does” - P7\\ “you have to think about your logic but it just does the work for you” - P2\\ “if a student uses code completion often in a lot of their assignments and work, they could find themselves relying on it a lot, not learning to manually code themselves” - P8\\ “Take advantage of it and make it finish their work for them instead of doing it themselves”- P10\\ “I think it’s more on their understanding of programming, of course we won’t let them rely too much on code completion to complete their work” - P9\end{tabular} \\ \hline
Academic Assessment Issues & 20\% & \begin{tabular}[c]{@{}p{0.75\textwidth}@{}}“It would be more important to manually code solutions for things like graded assignments as it reflects more on the skills of the student rather than their ability to use code completion.” - P8\\ “If you’ve got an autocomplete to help you write the algorithm, then you’re not proving that you know the algorithm, you’re just proving that you know how to use an autocomplete tool” - P6\end{tabular} \\ \hline
Skeptical \break Code Quality & 20\% & \begin{tabular}[c]{@{}p{0.75\textwidth}@{}}“if you provide a really high-level task then it probably won’t be able to help you. But some common things it is able to work well.” - P2\\ “remind them that code completion will not give 100\% accuracy for the code and to solve the problem. It’s there for them to guide them on how to solve the problems.” - P9\end{tabular} 
\end{tabular}
\end{adjustbox}
\end{table*}

\section{What are the students' challenges when adopting the AI code completion tool?}

\hspace{4mm}

\chapterinitial{``\emph{The over-reliance on AI code completion may lead to a surface-level understanding of programming concepts, diminishing problem-solving skills and restricting creativity.}''}

In addition to the benefits, students also raised several concerns when adopting AI code completion to write code, specifically, on the over-reliance on the suggestions. 
Here are some of the challenges that students encountered:

\textbf{Over-reliance on tools}: Nearly every student raises concerns about the risks associated with over-reliance on AI code completion tools, hindering students' learning and growth in coding and problem-solving.
In particular, when adopting AI code completion, such tools may lead to a surface-level understanding of programming concepts. 
Students might not fully grasp the underlying logic and syntax, as the tool completes the code for them. 
This can hinder their ability to apply their knowledge to new and complex problems.
For example, a participant expressed, \emph{``you have to think about your logic but it just does the work for you"}.
These findings highlight that the learning process may be impeded as AI code completion often accomplishes tasks without students comprehending the code. 


\textbf{Academic Assessment Issues}: It can be challenging for educators to assess a student's true coding abilities if they heavily rely on AI code completion tools. 
For instance, a participant expressed, \emph{``if you’ve got an autocomplete to help you write the algorithm, you’re just proving that you know how to use an autocomplete"}.
In addition, grading becomes more complex when it's unclear how much of the work is genuinely the student's own.
To address this challenge, several mitigation strategies are recommended including documentation (explanation of code's logic), oral examination (students discuss their code and explain their thought process), randomized assessments, individual interviews, etc.

\begin{table*}[t]
\centering
\caption{The expectations of using AI code completion from students' perspectives.}
\label{table2}
\begin{adjustbox}{width=\textwidth}
\begin{tabular}{p{0.15\textwidth}|c|p{0.75\textwidth}}
\hline
\textbf{Categories}& \textit{\textbf{\%}} & \textbf{Quotes} \\ \hline
Explaining Suggested Code & 30\% & \begin{tabular}[c]{@{}p{0.75\textwidth}@{}}“if it’s possible to have detailed comments for each line or every block of code which tries to explain what each block is doing.” - P8\\ “if we have a comment, between the lines for a better explanation that would be helpful for students to know what they’re doing and what they’re reading from code completion.” - P9\\ “Ability to hover over the top and showcase what it does”- P10\end{tabular} \\ \hline
Code Refactoring       & 30\% & \begin{tabular}[c]{@{}p{0.75\textwidth}@{}}“Highlight a certain part of the code, if the model can suggest a better version for that part of the code, would be more similar as to a teacher teaching” - P4\\ “Suggesting a better way to write something if it is inefficient or not following good coding practices, code completion could point that out” - P7\\ 
“if there is a more simple solution, that a simpler solution if possible is also achieved” - P8\end{tabular} \\ \hline
Specialized \break Code Completion       & 20\% & \begin{tabular}[c]{@{}p{0.75\textwidth}@{}}“Having an autocomplete model that has been purely trained for a specific use case. Jack of all trades, master of none concept.” - P6\\ “Optimise the code completion model to emphasize efficiency in coding as code suggested by code completion may not be most efficient” - P4\end{tabular}     \\ \hline
Personalized \break Code Completion     & 10\% & “more personalized. So some particular person's programming style. Like them suggesting edits will influence for them in particular. That's how I envision it.” - P3 \\ \hline
Natural Language \break Requirement Input  & 10\% & “Chat box to get the requirements, before giving suggestions as it might be more accurate to what the user wants” - P5 \\ \hline
Suggesting \break an Entire Solution   & 10\% & “It would be helpful if it can show a whole complete code for the question as it is quite time-consuming to click tab-by-tab for each line” - P9  \\ \hline
Suggesting \break Variable Names & 10\% & “variable names could be more representative of what they are being represented as in the code. If they were words or something that would be more helpful.” - P8     
\end{tabular}
\end{adjustbox}
\end{table*}

\textbf{Skeptical Code Quality}:
In principle, AI code completion is not specifically designed for generating high-quality code~\cite{liu2023refining}. 
Thus, AI code completion may not generate high-quality and accurate code solutions for every problem. 
For example, a participant stated that \emph{``remind them that code completion will not give 100\% accuracy"}.
This finding confirms that code quality issues become a central concern for students. 
Thus, students should only use AI code completion as a coding assistance, not an AI programmer.
Students should still be able to evaluate whether to accept or reject the suggestions by the AI code completion tool.

\section{What are the students' expectations when adopting the AI code completion tool?}

\hspace{4mm}

\chapterinitial{``\emph{In the future, AI code completion must be explainable to facilitate the learning of coding concepts.}''}

In addition to the benefits and challenges, students express many different expectations toward the improvement of adopting an AI code completion tool to write code. 
Here are some of the expectations from students:

\textbf{Explaining Suggested Code}: 
To enhance comprehension, students anticipate that AI code completion tools should have the capability to provide detailed explanations of the suggested code, whether through comments or hovering information.
For example, a participant emphasized, \emph{``if we have a comment between the lines for a better explanation that would be helpful for students"}.
This underscores the significance of providing explanations alongside suggested code to facilitate the learning of coding concepts, particularly for beginners.

\textbf{Code Refactoring}: 
Students anticipate that AI code completion can assist in recommending improved versions of their code.
Particularly in the initial learning stages, students might not be familiar with the best coding practices.
For instance, a participant highlighted, \emph{``Suggesting a better way to write something, if it is inefficient or not following good coding practices"}.
Therefore, this capability would function as a coding tutor, effectively aiding students in improving their code by recommending more efficient or better coding practices.

\textbf{Specialized/Personalized Code Completion}:
Students anticipate a code completion model optimized for solving specific tasks and enhancing the efficiency of code execution. 
As one participant mentioned, \emph{``Having an autocomplete model that has been purely trained for a specific use case"}.
Moreover, another participant desires a personalized code completion model capable of adapting to their individual coding style, referring to it as \emph{``more personalized''}.
Therefore, students hold the expectation of having AI code completion tailored to their unique needs and preferences.

\textbf{Other Expectations}:
From the students' perspective, other expectations include:
Incorporating natural language input to provide better context for AI code completion; 
AI code completion suggesting an entire solution at once, rather than suggesting line by line;
and providing suggestions for meaningful variable names.

\section{Conclusion}
The advent of AI code completion has fundamentally transformed the learning experience and interaction of computer science students with programming languages. 
Within this article, we studied the benefits, challenges, and student expectations when adopting AI code completion, utilizing our Visual Studio Code Extension, \emph{AutoAurora}.
Through an interview study of ten participants, our study unveils that, according to students' perceptions, AI code completion could enhance productivity and efficiency, operating as an effective coding tutor throughout the learning process.

However, a predominant concern arises regarding over-reliance on these tools, potentially impeding students' abilities to solve complex problems and presenting challenges in academic assessments.
This highlights the need for educators to thoughtfully consider these implications when evaluating student performance. 
As these tools continue to evolve, their integration into computer science education is inevitable; hence, it becomes crucial to strike a balance between automation and genuine learning.

Additionally, our research explores student expectations for the future of AI code completion, revealing that beginners may possess only a high-level understanding of the code suggested by the tools.
Consequently, the provision of explanations alongside the suggested code to enhance comprehension and refactoring versions to promote the adoption of best coding practices emerge as a pivotal component in the learning process.



\bibliographystyle{IEEEtranS}

\bibliography{mybibfile}

\begin{IEEEbiography}{Wannita Takerngsaksiri}
is a Ph.D. candidate at Monash University, Australia. 
Her research interests include code generation, machine learning (ML), and natural language processing (NLP). 
Specifically, her research aims to explore various aspects of code completion and to develop computational methods and advanced AI techniques. 
The goal is to assist the coding process, making it more intelligent and reliable for developers in practical applications.
Contact her at wannita.takerngsaksiri@monash.edu.
\end{IEEEbiography}

\begin{IEEEbiography}{Cleshan Warusavitarne} is pursuing a Bachelor's degree in Software Engineering (Honours) at Monash University, Australia.
Contact him at cwar0007@student.monash.edu.
\end{IEEEbiography}

\begin{IEEEbiography}{Christian Yaacoub} is pursuing a Bachelor's degree in Software Engineering (Honours) at Monash University, Australia. His hands-on experience includes developing custom software solutions for client projects and actively engaging in research projects. His specific focus lies in the integration of AI to tackle complex challenges and enhance user experience.
Contact him at christianyaacoub@gmail.com.
\end{IEEEbiography}

\begin{IEEEbiography}{Matthew Hee Keng Hou} is pursuing a Bachelor's degree in Software Engineering (Honours) at Monash University, Australia.
Contact him at mhee0019@student.monash.edu.
\end{IEEEbiography}

\begin{IEEEbiography}{Chakkrit Tantithamthavorn}
is a Senior Lecturer in Software Engineering and a 2020 ARC DECRA Fellow in the Faculty of Information Technology, Monash University, Australia. 
His research is focused on developing AI-enabled software development techniques and tools in order to help developers improve developers productivity, improve the quality of software systems, and make better data-informed decisions.
Contact him at chakkrit@monash.edu.
\end{IEEEbiography}

\end{document}